%%%%%%%%%%%%%%%%%%%%%%%%%%%%%%TEX-FILE%%%%%%%%%%%%%%%%%%%%%%%%%%%%%
%%%%%%%%%%%%%%%%%%%
\input phyzzx.tex
\hfuzz 20pt
\font\mybb=msbm10 at 12pt

\def\Bbb#1{\hbox{\mybb#1}}

\def\bR{\Bbb {R}}

\def\bfomega{\omega\kern-7.0pt \omega}

\def\C{\mkern1mu\raise2.2pt\hbox{$\scriptscriptstyle|$}\mkern-7mu{\rm C}}

%\sequentialequations

%%%%%%%%%%%%%%%%%%%%%%%%%%%%%%%%%%%%%%%%%%%%%%%%%%%%%%%%%%%%%%%%%%%%
\REF\aff{V. de Alfaro, S. Fubini and G. Furlan,  
{\sl Conformal Invariance in Quantum Mechanics}, 
Nuovo Cimento {\bf 34A} (1976) 569.}
\REF\akulov{V.P. Akulov and I.A. Pashnev, {\sl Quantum 
Superconformal Model
in (2,1) Space}, Theor. Math. Phys. {\bf 56} (1983) 862.}
\REF\fubini{ S. Fubini and E. Rabinovici, {\sl Superconformal 
Quantum Mechanics},
Nucl. Phys. {\bf B245} (1984) 17.}
\REF\ivanov{ E. Ivanov, S. Krivonos and V. Leviant, {\sl Geometry of
Conformal Mechanics}, J. Phys. {\bf A22} (1989) 345.}
\REF\freedman{D.Z. Freedman and P. Mende, {\sl An 
Exactly Solvable N Particle
System in Supersymmetric Quantum Mechanics},
 Nucl. Phys. {\bf B344} (1990) 317.}
\REF\claus{P. Claus, M. Derix, R. Kallosh, 
J. Kumar, P. Townsend and A. van

Proeyen, {\sl Black Holes and Superconformal Mechanics},
 Phys. Rev. Lett.
{\bf 81} (1998) 4553; hep-th/9804177.}
\REF\jade{J.A. de Azcarraga, J.M. Izquerido, 
J.C. Perez Buono and P.K. Townsend,
{\sl Superconformal Mechanics, Black Holes, and 
Non-linear Realizations}, Phys. Rev.
{\bf D59} (1999) 084015; hep-th/9810230.}
\REF\wyllard { N. Wyllard, {\sl (Super)Conformal 
Many Body Quantum 
Mechanics with Extended
Supersymmetry}; hep-th/9910160.}
\REF\maldacena{ J. Maldacena, {\sl The Large N Limit of 
Superconformal Field
Theories and Supergravity}, Adv. Theor. Math. Phys. 
{\bf 2} (1998) 231; hep-th/9711200.} 
\REF\gibbonst{ G. W. Gibbons and P.K. Townsend, {\sl 
 Black Holes and Calogero Models},  
 Phys.Lett. {\bf B454}(1999) 187; 
 hep-th/9812034.}
\REF\coles{ R. A. Coles and
 G. Papadopoulos, {\sl The Geometry 
of the One-dimensional
Supersymmetric Non-linear Sigma Models}, 
Class. Quantum Grav. {\bf 7} (1990)
427-438.}
\REF\gibbonsp{G.W. Gibbons, G. Papadopoulos  and K.S. Stelle, 
{\sl HKT and OKT Geometries on Soliton Black Hole Moduli Spaces},
 Nucl.Phys. {\bf B508} (1997)623; hep-th/9706207.}
\REF\michb{J. Michelson and A. Strominger, {\sl The 
geometry of (Super)conformal
Quantum Mechanics}, HUTP-99/A045, hep-th/9907191.} 
\REF\micha{J. Michelson and A. Strominger, {\sl Superconformal 
Multi-Black Hole
Quantum Mechanics}, HUTP-99/A047, hep-th/9908044.\break
R. Britto-Pacumio, J. Michelson, 
A. Strominger and A. Volovich,
{\sl Lectures on superconformal quantum 
mechanics and multi-black hole
moduli spaces}, hep-th/9911066.}
\REF\mss{A. Maloney, M. Spradlin and
 A. Strominger, {\sl Superconformal
Multi-Black Hole Moduli Spaces 
in Four Dimensions}, hep-th/9911001.}
\REF\gibr {G.W.Gibbons \& P.J.Ruback,{\sl
The motion of extreme Reissner-Nordstr\"om black
holes in the low velocity limit}, Phys. Rev.
Lett. {\bf 57} (1986) 1492.}
\REF\fere {R.C.Ferrell \& D.M.Eardley,{\sl
Slow motion scattering and coalescence of
maximally charged black holes},Phys. Rev.
Lett. {\bf 59} (1987) 1617.}
\REF\shiraishi{K. Shiraishi, {\sl Moduli Space 
Metric for Maximally-Charged
Dilaton Black Holes}, Nucl. Phys. {\bf B402} (1993) 399.}
\REF\gutpap{J. Gutowski and G. Papadopoulos, {\sl The dynamics 
of very special black holes}, hep-th/9910022 .}
\REF\moore{R. Minasian, G. Moore, D. Tsimpis,
 {\sl Calabi-Yau Black Holes and 
 (0,4) Sigma Models}; hep-th/9904217.}
\REF\hellerman{S. Hellerman and 
J. Polchinski, {\sl Supersymmetric 
Quantum Mechanics from Light Cone Quantization}; hep-th/9908202.}
\REF\hull {C.M. Hull, {\sl The Geometry of 
Supersymmetric Quantum Mechanics};
 hep-th/9910028.} 
\REF\hullp {C.M. Hull, G. Papadopoulos and P.K. Townsend, 
{\sl Potentials for (p,0) and (1,1) Supersymmetric
 Sigma Models with Torsion},  Phys.Lett. {\bf B316} (1993) 
 291; hep-th/9307013.} 
\REF\gp{G. Papadopoulos, {\sl Topology, Unitary 
Representations and Charged
Particles}, Commun. Math. Phys. (1992) 144.} 
 \REF\witten{C.M. Hull and  E. Witten,
 {\sl Supersymmetric Sigma Models and the Heterotic String}, 
  Phys.Lett. {\bf B160} (1985) 398.} 
  \REF\hpa{P.S. Howe and G. Papadopoulos,
{\sl Further Remarks on the Geometry of Two-dimensional 
Nonlinear Sigma
Models}, Class.Quant.Grav. {\bf 5} (1988) 1647.} 
\REF\vn{G. W. Delius, M. Rocek, A. Sevrin and 
P. van Nieuwenhuizen,
 {\sl Supersymmetric Sigma Models with 
 Non-vanishing Nijenhuis Tensor and
their Operator Product Expansion},
 Nucl.Phys. {\bf B324} (1989) 523. }
 \REF\hpb{P.S. Howe and G. Papadopoulos, {\sl  Holonomy 
 Groups and W Symmetries},
 Commun.Math.Phys. {\bf 151} (1993) 467; hep-th/9202036.}
 \REF\gpb{G. Papadopoulos, {\sl (2,0)-Supersymmetric Sigma 
Models and Almost Complex Structures},
 Nucl. Phys. {\bf B448} (1995) 199;
 hep-th/9503063.}
\REF\sfetsos{F. De Jonghe, K. Peeters  and K. Sfetsos,
 {\sl Killing-Yano Supersymmetry in String Theory},
 Class.Quant.Grav. {\bf 14} (1997) 35;  hep-th/9607203.}   
\REF\twist{P.S. Howe and G. Papadopoulos, {\sl Twistor Spaces
for HKT Manifolds}, Phys. Lett. 
{\bf B379} (1996)80, hep-th/9602108.}

%%%%%%%%%%%%%%%%%%%%%%%%%%%%TITLE PAGE%%%%%%%%%%%%%%%%%%%%%%%%%
\Pubnum{ \vbox{ \hbox{}\hbox{} } }
\pubtype{}
\date{February, 2000}
\titlepage
\title{Conformal and Superconformal Mechanics}
\author{ G. Papadopoulos }
\address{Department of Mathematics\break King's College London \break  
Strand\break London WC2R 2LS}
%\andauthor{}
%\address{}
%\address{}
\abstract {We investigate the 
conformal and superconformal properties
of a non-relativistic spinning 
particle propagating in a curved background
coupled to a magnetic field and with a scalar potential.
We derive the conditions on the 
couplings for a large class of such systems
which are necessary in order their 
actions  admit conformal 
and superconformal symmetry. We find 
that some of these conditions
can be encoded in the conformal and 
holomorphic geometry of the background.
Several
new examples of conformal and 
superconformal models are also given.}

\endpage
\pagenumber=2

%%%%%%%%%%%%%%%%%%%%%%%%%%%%%%%%%%%%%%%%%%%%%%%%%%%%%%%%%%%%%%%%%%

%%%%%%%%%%%%%%%%%%%%%%%%%%%%%%%%%%%%%%%%%%%%%%%%%%

\chapter{Introduction}

Conformal  transformations  leave  the metric 
of a manifold invariant up to a coordinate dependent scale. 
Many   theories exhibit conformal invariance and the classical
and quantum consequences of this have been extensively explored
 in the literature. A celebrated application 
is the use 
of  superconformal two-dimensional
sigma models 
to describe the consistent  propagation of superstrings in 
curved backgrounds.

More recently, there has been much 
interest in classical and quantum 
mechanical systems which exhibit 
conformal  symmetry following
some earlier work by de Alfaro, 
Fubini and Furlan [\aff]. Superconformal
models have been examined in [\akulov-\wyllard].
Such systems are sigma models that 
describe the propagation of a non-relativistic
spinning particle  in a curved 
background which is the  sigma
model target space. 
  Conformal sigma models may have applications in the 
  context of AdS/CFT correspondence
adapted for  $AdS_2\times M$ backgrounds 
[\maldacena]. Another application
for these models is in the study of 
the radial motion of non-relativistic
 test particles
with non-vanishing angular 
momentum near the horizon of extremal Reissner-
Nordstr\"om black holes 
[\claus, \jade]. It turns out 
that the Hamiltonians of the
test particles are either given 
by that of [\aff] or by a suitable
 generalization [\gibbonst]; the target spaces of all these models
are flat but they have {\it scalar} potentials. In addition, 
supersymmetric [\coles, \gibbonsp] and  
superconformal [\michb-\mss] 
one-dimensional sigma
models have found applications in the study of the moduli spaces
of  black holes in four and five 
dimensions [\gibr-\shiraishi, \gibbonsp, \michb-\mss, \gutpap] 
that preserve a fraction
of spacetime supersymmetry.
In particular, it has been shown in 
[\michb-\mss] that for a certain class
of black holes the effective theory is 
described by a supersymmetric
sigma model which at small black hole 
separations it exhibits
superconformal symmetry. In this case, the 
target space of the relevant
 sigma model
is curved and there is no scalar potential. 
A different application
of (two-dimensional) sigma model is in the 
context of the moduli space
of  Calabi-Yau black holes
 [\moore]. Some other  applications of
one-dimensional supersymmetric sigma models have been in the 
light-cone quantization of supersymmetric theories
[\hellerman].

It has been known for sometime that there is an interplay between
the number of supersymmetries of sigma models and the geometry
of their target spaces. The supersymmetric one-dimensional
sigma models have been investigated in [\coles].   It was found that
 supersymmetry in this case imposes weaker conditions
on the geometry on the target spaces than those imposed
on the target spaces of  sigma models with the 
same number of supersymmetries but in two or more dimensions.
It turns out that in one-dimension there is some flexibility in
 the derivation of the various conditions that supersymmetry
imposes on the couplings of the theory. This makes the analysis
of the conditions rather involved and a complete
classification of all possibilities has not been done.
Some new one-dimensional supersymmetric sigma models have 
recently been constructed in  [\hull].

Superconformal symmetry imposes further conditions in addition
to those of supersymmetry  on the
target space of one-dimensional sigma models.
These additional conditions were first explored in [\michb, \mss] for
a class of supersymmetric sigma models which did not have
 a coupling to a magnetic field or scalar potential. In particular
it was found that a necessary condition for a sigma model
to have superconformal symmetry is that the sigma model
manifold   admits a homothetic
motion.

In this paper, we investigate the conditions for a 
non-relativistic spinning particle coupled to a {\sl magnetic
field}, with a {\sl scalar potential} and propagating in a curved
background to admit conformal and  superconformal symmetry.
We shall carry out our analysis in the Lagrangian formalism.
We first begin with the bosonic spinning particle.
The conformal group of the real line is
 the group of its diffeomorphisms. Such transformations do not leave the
 non-relativistic spinning particle action invariant unless they are 
 appropriately compensated 
 with diffeomorphisms of the background.
We derive the conditions that the background should satisfy in order 
to admit such diffeomorphisms which induce conformal symmetries in the
 action of the spinning particle. In particular 
we find that the
necessary conditions for the spinning particle 
action to be invariant under the
$SL(2,\bR)$ subgroup of the conformal group are 
that either the sigma model
manifold admits a homothetic motion or it admits two 
commuting homothetic motions.
We apply our results to the case of a charged 
particle in the presence
of a Dirac monopole and to the propagation of a particle in a conical
space.
We also give the various
 conserved charges and compute their poisson bracket algebra.
 We then extend our analysis to the 
superconformal case. For this
we consider a class of supersymmetric 
non-relativistic spinning particles
coupled to a magnetic field and with a scalar potential.
To incorporate the latter coupling into the action, we include 
a fermionic multiplet in analogy with similar results in
two-dimensional sigma models  [\hullp]. The scalar potential 
is the length of  section of a vector 
bundle over the background manifold.
We first give the conditions for the 
$N=1$ supersymmetric
models to admit a superconformal extension. 
In particular, we investigate
the conditions for the models to be 
invariant under the $Osp(1|2)$ subgroup
of the one-dimensional $N=1$ 
superconformal group. The conditions for $Osp(1|2)$
invariance are closely related to 
those of the $SL(2,\bR)$ invariant bosonic models.
We give some examples of $Osp(1|2)$ invariant models. 
Next we extend our analysis to a class of $N=2B$ and $N=4B$ 
supersymmetric one-dimensional
systems with a magnetic field 
and scalar potential. In particular 
we derive the conditions
for such models to be invariant 
under the $SU(1,1|1)$ and $D(2,1|\alpha)$ 
subgroups of the $N=2B$ and $N=4B$ superconformal groups in 
one-dimension, respectively. In the 
special case of sigma models without fermionic
superfields and without a coupling to a magnetic field 
the above conditions are related to
those in [\michb].

This paper has been organized as follows: 
In section two, we derive the
conditions for conformal invariance 
of a bosonic spinning particle.
In section three, we give the 
conserved charges and compute their
poisson bracket algebra. In section four, we investigate the conditions
for a $N=1$ supersymmetric spinning 
particle to admit superconformal symmetry.
In section five, we give the conditions 
for a $N=2B$ supersymmetric spinning particle
to admit  superconformal symmetry. In section 
six, we investigate the conditions
for a $N=4B$ supersymmetric spinning 
particle to admit superconformal
symmetry  and in section
seven we give our conclusions. In appendix A, 
we present a geometric
interpretation for the conformal 
transformations that we are considering.
In appendix B, we give the most 
general action for a $N=1$ supersymmetric
sigma model with  couplings which 
are at most quadratic in the velocities
of the bosonic fields.

\chapter{Conformal Bosonic Sigma Models}
\section{Conformal Invariance}

The conformal group, ${\rm Conf}(\bR^n)$ of 
n-dimensional Euclidean spaces, $\bR^n$, for $n\geq 3$, 
is a finite dimensional
Lie group which is isomorphic to  
$O(n+1,1)$. But in 
one- and two-dimensional Euclidean
spaces,  the conformal group is 
infinite dimensional. In one-dimension
the conformal group ${\rm Conf}(\bR)$ is isomorphic to the
diffeomorphism group, ${\rm Diff}(\bR)$ of the real line.    
So the Lie algebra ${\rm conf}(\bR)$  
of ${\rm Conf}(\bR)$ is the set
of vector fields
on $\bR$ equipped with the Lie bracket.  
The set of vector fields on $\bR$ 
can be identified  with the set of real functions
on $\bR$, and so ${\rm conf}(\bR)=C^\infty (\bR)$. 
In this identification
a vector field on $\bR$ is associated 
to its component in $C^\infty (\bR)$.
So, it is not realistic to expect that 
we shall be able to construct
models which   are invariant under the
full conformal group in one-dimension;
invariant models under the full conformal group will be 
necessarily diffeomorhic invariant
and so a coupling to gravity may have to be included.
Therefore of interest are the models 
which are invariant under a proper subgroup
of the conformal group in one-dimension.
One such case is the
subalgebra of ${\rm conf}(\bR)$ generated by the
polynomials $P[t]$ on the real line. 
This case includes the dilations $L$ and 
the special conformal transformations $T$  which have been
examined in the context of one-dimensional 
sigma models in [\michb].  The  translation $H$ 
along the worldline
 together with $L$ and $T$ generate the $sl(2,\bR)$ subalgebra,
$$
[L,T]=T,\qquad [H,L]=H, \qquad [H,T]={1\over2} L\  ,
\eqn\confsub
$$
of the conformal algebra in one-dimension.
The systems that are invariant under the 
subalgebra \confsub\ of the
conformal algebra of the real line were 
called conformal or conformally invariant.
However in what follows we shall not make 
this restriction. We shall investigate
conformal invariance under the full 
conformal group of the real line or a subgroup
of it that contains \confsub.
As we shall see though many interesting 
systems are invariant only under
the action of the subalgebra \confsub.

Our approach to investigate the conformal symmetries
of the spinning particle is geometric, so
it is more convenient to use the sigma 
model language to describe the
various computations.  Let $M$ be 
a Riemannian manifold equipped with
a metric $g$, a locally defined 
(up to an exact one-form) one-form $A$ and a locally defined 
(up to a constant) function $V$.  The Lagrangian 
of
a one-dimensional sigma model with target space $M$ is 
$$ 
L={1\over2} g_{ij} {d\over dt} q^i  {d\over dt} q^j 
+ A_i  {d\over dt} q^i-V(q)\ ,
\eqn\blag 
$$ 
where $t$ is a coordinate on the worldline $\bR$ 
and $q$ are the sigma model maps from $\bR$ into $M$.
The couplings $A$ and $V$ are those of 
a magnetic field and a scalar
potential, respectively. Such Lagrangian  
describes a non-relativistic
spinning particle coupled to a magnetic 
field $A$ with scalar potential $V$
and propagating in the background with metric $g$.
The field equations are
$$
\nabla_t{d\over dt} q^i-g^{ik} F_{kj} 
{d\over dt} q^j +g^{ij}\partial_j V=0\ ,
\eqn\mone
$$
where $F_{ij}= 2 \partial_{[i} A{_j]}$. Note that the field
equations are expressed in terms of globally defined tensors on $M$.

The action of the sigma model \blag\ is 
invariant under worldline translations
$\delta t=\epsilon$ which are generated by
$a=d/dt$. The rest of the conformal transformations
 do {\sl not} leave \blag\ invariant. 
 However we can circumvent
this by introducing a 
 transformation for $q$
which is generated by a vector field $X$ in 
$M$. We also allow $X$ to depend
explicitly on $t$. The vector fields $X$ generate 
an one-parameter family of
diffeomorphism of $M$ and  generic diffeomorphisms
do {\sl not} leave the sigma model
action invariant, i.e. there are not conserved  charges associated
with them.   But if a conformal transformation is appropriately
compensated with a diffeomorphism of $M$, then  under certain 
conditions their 
combination can
 leave the action invariant. So we  seek the 
 conditions for the Lagrangian
\blag\ to be invariant up to possibly surface terms
 under the transformations\foot{The geometric interpretation
 of these transformations is given in appendix A.}
$$
\eqalign{
{\bar \delta_\epsilon}q^i&=\epsilon X^i(t, q)
\cr
{\bar \delta_\epsilon}t&=\epsilon a(t)\ ,}
\eqn\mtwo
$$
where $\epsilon$ is a constant infinitesimal parameter.
The induced transformation on $q$ is
$$
\delta_\epsilon q^i = - \epsilon a(t) 
{d\over dt} q^i +\epsilon X^i(t, q)\ .
\eqn\btran
$$
To continue we note that if $f=f(t, q(t))$, then the total
derivative ${d\over dt}$ of $f$ with respect to $t$ is 
$$
{d\over dt}f= \partial_t f+\partial_i f {d\over dt}q^i\ ,
\eqn\mthree
$$
where   $\partial_t$ denotes differentiation which acts only
on the explicit dependence of $f$ on  $t$.
For functions that depend {\sl only} on $t$, like
$q$ and $a$, ${d\over dt}$ is the same as $\partial_t$, and
in what follows  we shall use either
 one or the other notation at convenience.  
The commutator of two such transformations yields
$$
[\delta_\epsilon, \delta_{\epsilon'}] q^i
=-\epsilon \epsilon' [a, a']^t  
{d\over dt} q^i +\epsilon \epsilon' \big( a{ \partial_t} X'^i-
a'{\partial_t} X^i+ [X,X']^i\big)\ ,
\eqn\mfour
$$
where $[a, a']$ and $[X,X']$ are the Lie 
brackets of the vector fields $a, a'$ and
$X,X'$, respectively. The commutator is a 
new transformation on $q$ induced by
the Lie bracket of the vector fields 
$a, a'$ as expected and by a new 
vector field $Y$
on $M$ with components
$$
Y^i=a{\partial_t} X'^i-
a'{ \partial_t} X^i+ [X,X']^i\ .
\eqn\mfive
$$
Observe that this new vector field 
is of the same type as that of $X$ and $X'$
and so the algebra of transformations closes.
The Lagrangian \blag\ is invariant under the 
transformations \btran\ provided
that
$$
\eqalign{
\nabla_{(i} X_{j)}&= {1\over2} \partial_t a g_{ij}
\cr
{\partial_t} X^j g_{ji} + X^j F_{ji}&= \partial_i f
\cr
\partial_t a V + X^i \partial_i V&=-\partial_t f\ ,}
\eqn\binv
$$
where $f=f(t, q)$ is a function on $\bR\times M$. The
function $f$  arises
because the Lagrangian \blag\ should be invariant
up to a surface term. In general $f$ could be chosen to depend
on the time derivatives of $q$ as 
well, however, since our Lagrangian
is at most quadratic in the time derivatives of 
the fields the above choice of $f$ suffices. 

\section{Special Cases}

Next we consider the class of conformal 
transformations generated by
the polynomials $P[t]$ on $\bR$. We also 
assume that the vector fields
$X$ and the boundary term $f$ are polynomials in $t$. So we have
$$
\eqalign{
a(t)&= \sum_{n=0}^I a_n t^n
\cr
X(t, q)&=\sum_{n=0}^J X_n(q) t^n
\cr
f(t, q)&=\sum_{n=0}^K f_n(q) t^n}\ .
\eqn\aone
$$
Substituting these into the equations 
for the invariance of the
action we find that
$$
J\geq I+1
\eqn\atwo
$$
and
$$
\eqalign{
\nabla_{(i} X_{n j)}&= {1\over2} (n+1) a_{n+1} g_{ij}
\cr
(n+1) X_{n+1}^j g_{ji} + X_n^j F_{ji}&= \partial_i f_n
\cr
(n+1) a_{n+1} V + X_n^i \partial_i V&=-(n+1)f_{n+1} \ ,}
\eqn\binvn
$$
For example, if we take
$$
\eqalign{
a&=t
\cr
X(t, q)&=Z(q)}
\eqn\athree
$$
which are the dilations in $\bR$, we find that
$$
\eqalign{
\nabla_{(i} Z_{j)}&= {1\over2}   g_{ij}
\cr
 Z^j F_{ji}&= \partial_i f(q)
\cr
 V + Z^i \partial_i V&=0\ ,}
\eqn\bdinv
$$
up to a possibly constant shift of $V$.
So $Z$ generates a homothetic motion on $M$ which leaves
$F$ invariant. 
For special conformal transformations 
$$
a=t^2\ .
\eqn\afour
$$
A  choice for  $X$ is
$$
X(t,q)=tY(q)\ .
\eqn\afive
$$
Substituting these into \binvn, we  find
$$
\eqalign{
\nabla_{(i} Y_{ j)}&=  g_{ij}
\cr
 Y^j g_{ji}&=\partial_i f_0
 \cr
 Y^j F_{ji}&= \partial_i f_1
\cr
2 V + Y^i \partial_i V&=-f_1\ ,}
\eqn\bsinv
$$
So $Y$ is  a homothetic vector field 
on $M$ which leaves $F$ invariant
and is generated by a \lq homothetic potential' $f_0$. 
We can choose the vector field $Y$ above to be different from $Z$ 
in \bdinv. But a minimal choice is to set 
$$
Y=2Z \ .
\eqn\minm
$$
 In such which case,
$f_1=0$ and we find that
$$
\eqalign{
\nabla_{(i} Z_{ j)}&= {1\over2} g_{ij}
\cr
 Z^j g_{ji}&={1\over2}\partial_i f_0
 \cr
 Z^j F_{ji}&= 0
\cr
V + Z^i \partial_i V&=0\ .}
\eqn\bsinvc
$$  
These are the conditions for a sigma model to be invariant under
the dilations and special conformal transformations 
generated by a single dilatation $Z$ in the target space $M$.
In particular, observe that
$$
Z^j g_{ji}={1\over2} \partial_i f_0
\eqn\kpot
$$
where $f_0$ is  a homothetic potential.
No further conditions are required by the closure of the algebra
of transformations. 
For $F=0$, this
reproduces the results of [\michb]. If  $M=\bR^d$ and $F=0$, 
then $X={1\over2} q^i\partial_i$ and $V$ scales with degree $-2$. For
$d=1$, this gives $V=q^{-2}$
which is the standard potential of conformal mechanics [\aff].

Next let us consider the invariance of the 
action under the transformation
generated by 
$$
\eqalign{
a&=t^{n+1}
\cr
X&= (n+1) t^{n} Z}
\eqn\ninva
$$ 
for $n\geq 2$, where $Z$ generates a homothetic motion.
Clearly the first condition in \binvn\ is satisfied. 
The remaining conditions
are
$$
\eqalign{
(n+1) n Z^j g_{ji}&=\partial_i f_{n-1}
\cr
(n+1) Z^J F_{ji}&=\partial_i f_n
\cr
\partial_i f_k&=0, \qquad k\not= n , n-1
\cr
 V+ Z^j \partial_j V&=- f_{n+1}
 \cr
f_k&=0, \qquad k\not= n+1  }
\eqn\asix
$$
Now since $V+ Z^j \partial_j V=0$, \asix\ implies that $f=0$. 
So we conclude that
$$
Z^j g_{ji}=0
\eqn\aseven
$$
and if the metric is non-degenerate, then $Z=0$.
So there are no additional such 
symmetries generated by a
single homothetic motion in $M$ apart
from those of dilations and special
 conformal transformations above.

Apart from the minimal case we have considered above,
we can also choose  $Z$ and $Y$ in \bdinv\ and in \bsinv,
respectively, to be  linearly independent.
Both $Z$ and $Y$ generate homothetic motions on $M$. 
It is straightforward to observe from \bdinv\ and \bsinv\ that
although $Y$ is associated to a homothetic potential, $Z$ does not.
The algebra of the transformations  generated by $Z$ and $Y$ closes
without the addition of other transformations
provided that 
$$
[Z,Y]=0\ .
\eqn\comcom
$$ 
The algebra of transformations is given by 
\confsub. One can easily extend the above results to the case that
the sigma model manifold admits a homothetic group 
action which generates more
than two vector fields on $M$. 

To summarize, the necessary conditions for a 
spinning particle propagating in a curved
background to admit a $sl(2,\bR)$ are either 

the existence of a homothetic motion
in the background generated by a homothetic 
potential or the presence of two commuting
homothetic motions. Additional conditions 
though are satisfied by the other couplings
of the theory like that of the magnetic field 
and scalar potential.

\section{Examples and Applications}

\subsection{Dirac Monopole}

As an example, we consider a non-relativistic particle in $\bR^3$
coupled to a Dirac monopole located at the origin.
We take $Z={1\over2} q^i\partial_i$. Clearly the first condition
in \bsinvc\ is satisfied ($g_{ij}=\delta_{ij}$). Choosing 
$f_0={1\over2} |q|^2$,
we can easily verify the second condition 
in \bsinvc. Moreover since
$$
F_{ij}=- {\epsilon_{ijk} q^k\over |q|^3}
\eqn\aeight
$$
we find that $ Z^jF_{ji}=0$ and the third 
condition in \bsinvc\ is also
satisfied. Finally the potential $V$ can  either vanish or be any
homogeneous function of $q$ of degree $-2$.
Therefore the system is invariant under dilatations and the special
conformal transformations on $\bR$ which together 
with the translations in $\bR$
generate an $sl(2,\bR)$ symmetry. Apart from these
symmetries the action is also invariant 
under $so(3)$ rotations provided 
the potential either vanishes or $V={1\over |q|^2}$.
These are generated by 
$X_i=\epsilon_{ijk} q^j \partial_k$ and $a=0$. One can easily
show that $[Z, X_i]=0$. As a result 
the transformations generated by the
rotations commute with those of $sl(2,\bR)$. 
The full symmetry of the system
is $so(3)\oplus sl(2,\bR)$  

\subsection{Particles propagating in Cones}

A large class of manifolds $M$ that admit a 
homothetic motion are those
that are cones over another manifold $N$, 
i.e. $M=C(N)$ and the metric of $M$
can be written as
$$
ds^2(M)=dr^2+r^2 ds^2(N)\ ,
\eqn\cmetric
$$
where $r$ is a radial coordinate. This example 
has also been considered in
[\michb]. The homothetic
vector field is
$$
Z=r {\partial\over \partial r}\ .
\eqn\anine
$$

Conical spacetimes have found applications in 
the context of M- and string
theories. It is well known that there are
 M- and string brane solutions
for which near the horizon are $AdS_d\times N$
 and at infinity
are $\bR^{(1, d-2)}\times C(N)$. The manifold 
$N$ is usually chosen to
be a coset space $G/H$ which occurs in 
various compactifications of
eleven- and ten-dimensional supergravity theories.

Next consider a line bundle over $N$ with connection $A$ which we can
use as a magnetic field. It is
straightforward to observe that all the conditions required 
for conformal invariance of the associated sigma model are met
provided that we choose
$$
V=r^{-2} U\ ,
\eqn\aten
$$
where $U$ is a function on $N$.

Next if  $N=G/H$ and we require invariance of 
the spinning particle action under $G$,  the
metric, magnetic field and scalar potential should be invariant
under $G$. In particular this implies that 
the function $U$ in the potential $V$
should be  constant. The symmetries generated 
by $G$ and those of $Sl(2,\bR)$ commute.
The symmetry of the system is $SL(2,\bR)\times G$.

\chapter { Conserved Charges}
\section{Poisson Bracket Algebra}
The conformal transformations 
investigated in the previous section
leave the sigma model action invariant up to surface terms.
Taking these surface term into 
consideration, we find that 
the conserved  charges  associated 
with the \btran\ transformations
 are
$$
C(a,X)=-{1\over2} a g_{ij} \partial_t q^i  
\partial_t q^j + g_{ij} \partial_t q^i X^j -a V(q) -f \ .
\eqn\aeleven
$$
It is straightforward to show after some computation
 that they are conserved subject to field equations using \binv.
 These charges are well-defined provided that $V$ and $f$ are
 functions on $M$ and  $\bR\times M$, respectively. 
The above charges are not
uniquely defined but they can shift up to a constant. This
is because $f$ is defined up to a constant.

The algebra of charges is usually computed by turning into
Hamiltonian formalism. However, this need not be the case
because a Poisson-like bracket can be defined within the Lagrangian
approach\foot{This way of defining Poisson brackets is also known as
the covariant canonical approach to classical mechanics.} as
$$
\{C(a,X),C(a',X')\}\equiv \delta_\epsilon C(a',X')|_{\epsilon=1}-
\delta_{\epsilon'} C(a,X)|_{\epsilon'=1}\ . 
\eqn\bone
$$
For our conserved charges, we find that
$$
\eqalign{
\{C(a,X),C(a',X')\}&=2\big[-{1\over2} [a,a'] g_{ij} 
\partial_t q^i \partial_t q^j
\cr
+&\big( \partial_t X'^j-a' \partial_t X^j +[X,X']^j \big)
 g_{ji} \partial_t q^i-[a,a'] V(q)
\cr
-&(\partial_tf' a-\partial_tf a'+F_{ij} X^i X'^j
+X^j\partial_jf'-X'^j \partial_j f)\big]
\cr
& -a g_{ij} {\cal S}^i X'^j+a' g_{ij}{\cal S}^i X^j\ , }
\eqn\btwo
$$
where ${\cal S}$ are the equations of motion.
Neglecting the terms involving the field
equations, this algebra can be rewritten  as
$$
\{C(a,X),C(a',X')\}=2 C([a,a'], [X,X'])+
 \kappa\big((a,X),(a',X')\big)\ ,
\eqn\bthree 
$$
where 
$$
\kappa\big((a,X),(a',X')\big)= f([a,a'], [X,X'])-(\partial_tf' 
a-\partial_tf a'+F_{ij} X^i X'^j
+X^j\partial_jf'-X'^j \partial_j f)
\eqn\bfour
$$
is a constant and $f([a,a'], [X,X'])$ is the surface term associated
with the commutator of the symmetries generated by 
$(a,X)$ and $(a',X')$. If $\kappa$ does not vanish, then the algebra
of charges may develop a central extension. This  
extension occurs if it cannot
be removed by shifting the charges up to 
constants. The central extension
defines an element in the second 
Lie-algebra cohomology of the symmetry group
and vanishes whenever the class is trivial.

\section{Examples}

In the special case for which $a=t^{n+1}$ 
and $X=(n+1) t^n Z$, $n=-1$ (translations),
$n=0$ (dilations) and $n=1$ (special conformal transformations)
investigated in  section (2.2), the conserved charges are
$$
C_{n+1}=-{1\over2} t^{n+1} g_{ij} \partial_t q^i  
\partial_t q^j +  (n+1)t^n g_{ij} \partial_t q^i Z^j -t^{n+1} V(q) -f \ .
\eqn\bfive
$$
In particular, we find that
$$
\eqalign{
C_0&=-{1\over2}g_{ij} \partial_t q^i  \partial_t q^j- V(q)
\cr
C_1&=-{1\over2} t g_{ij} \partial_t q^i  
\partial_t q^j +   g_{ij} \partial_t q^i Z^j -t V(q)
\cr
 C_2&=-{1\over2} t^2 g_{ij} \partial_t q^i  
\partial_t q^j +  2t  g_{ij} \partial_t q^i Z^j -t^2 V(q) -f_0\ ,}
\eqn\bsix
$$
where $f_0$ is given in \bsinvc.
The algebra of the conserved charges 
does not develop a central extension
because $sl(2, \bR)$ is semisimple.
The conserved charges of a particle 
in $\bR^3$ coupled to a Dirac
monopole are
$$
\eqalign{
C_0&=-{1\over2}\delta_{ij} \partial_t q^i  \partial_t q^j- V(q)
\cr
C_1&=-{1\over2} t \delta_{ij} \partial_t q^i  
\partial_t q^j +   \delta_{ij} \partial_t q^i Z^j -t V(q)
\cr
 C_2&=-{1\over2} t^2 \delta_{ij} \partial_t q^i  
\partial_t q^j +  2t  \delta_{ij} \partial_t q^i Z^j -t^2 V(q) 
- {1\over 2} |q|^2\ .}
\eqn\dcona
$$
The conserved charges of the 
rotational symmetries generated by  
$X_{(i)}=\epsilon_{ijk} q^j \partial_k$ are
$$
C_i=\epsilon_{ijk} q^j \partial_t q^k+ q_i |q|^{-1}\ .
\eqn\rotcharg
$$
The second term in the right-hand-side of \rotcharg\ is present 
because
$$
X^j_{(i)} F_{jk}= -\partial_k (q_i |q|^{-1})\ .
\eqn\dconb
$$
The algebra of the charges \dcona\ and \dconb\ does not
develop a central extension because $so(3)\oplus sl(2,\bR)$ is
semi-simple.

It is straight forward to give the charges of the conformal symmetries
of a spinning particle propagating in a 
conical space. For conical spaces
over $N=G/H$, the algebra of charges of
 the system is $sl(2,\bR)\oplus {\rm Lie} G$
provided that $G$ is semisimple. If
 $G$ is not semisimple and the charges
associated with $G$-invariance are
 well defined, there is the possibility
of a central extension in the algebra 
of charges of associated with 
$G$ symmetries. This is  due to the 
presence of the magnetic coupling.
For more details see for example [\gp].

\chapter{N=1 Superconformal Sigma Models}

\section{Superconformal Transformations}

In analogy with the bosonic models, we may 
identify the $N=1$ superconformal
algebra with the superdiffeomorphisms of the $N=1$
superspace $\Xi$. Let 
$(t, \theta)$ be   the commuting 
and anti-commuting coordinates of $\Xi$, respectively.
We can construct the even and odd vector fields on $\Xi$ as
$$
\eqalign{S_e&=a(t)\partial_t+ b(t) \theta\partial_\theta
\cr
S_o&= c(t) \partial_\theta +e(t) \theta \partial_t\ , }
\eqn\cone
$$
respectively, where $a,b,c,e$ are functions of $t$.

As we have already mentioned in the introduction, in many 
applications one considers the
subalgebra $Osp(1|2)$ of  vector fields on $\Xi$ with 
non-vanishing (anti-)commutators
$$
\eqalign{
\{Q,Q\}&=2H,\qquad [Q,L]=Q,\qquad \{S,Q\}=L,\qquad \{S, S\}=2T
\cr
[L,T]&=T,\qquad [L, S]= 2S,\qquad [Q,T]={1\over2} S,
 \qquad [S,H]=-{1\over2} Q
\cr
[H,L]&=H, \qquad [H,T]={1\over2} L, \qquad [L,T]=T}
\eqn\supconf
$$
where $H$ is the generator of translation, $L$ is the generator of
dilation, $T$ is the generator of special conformal, $Q$ is the 
generator of supersymmetry
 and $S$ is
the generator of special superconformal transformations. 
In terms of vector fields on $\Xi$ this subalgebra
 is realized\foot{We take the even and odd
coordinates of $\Xi$ to be real. Moreover 
we define the complex conjugation
of fermion bi-linears as 
$(\lambda_1\lambda_2)^*=\lambda_1^*\lambda_2^*$. 
In this notation 
bi-linears of real fermions are real without
 the use of the  imaginary unit; compare with the
conventional notation of e.g.[\coles].}
as follows:
$$
\eqalign{
H&=\partial_t,\qquad  Q=\partial_\theta+\theta \partial_t\ ,
\cr
L&=t\partial_t+{1\over2} \theta \partial_\theta,\qquad S={t\over2} Q,
 \qquad T={t^2\over4}
\partial_t+{t\over4}\theta \partial_\theta \ .}
\eqn\ctwo
$$

Next we introduce the $N=1$ superfields $q$ as functions from 
the superspace $\Xi$ into the sigma model manifold $M$.
The transformations induced on  superfields $q$ by the odd
vector fields are
$$
\bar\delta q^i=-\alpha \big[c(t) {\partial \over \partial\theta} q 
+e(t)\theta {\partial \over \partial t}\big] q^i\ ,
\eqn\cthree
$$
where $\alpha$ is an anti-commuting infinitesimal parameter.
Let $D={\partial\over\partial\theta}-\theta \partial_t$ be the 
supersymmetry derivative, i.e.
$$
QD+DQ=0, \qquad D^2=-\partial_t\ .
\eqn\cfour
$$
In components, the above transformations read
$$
\eqalign{
\bar\delta q^i&=-\eta c(t) \lambda^i 
\cr
\bar\delta\lambda^i&=+\eta e(t) \partial_t q^i\ ,}
\eqn\trana
$$
where
$$
\eqalign{
q^i&=q^i|
\cr
\lambda^i&=Dq^i|}
\eqn\cfive
$$
and the vertical line denotes evaluation at $\theta=0$. 
Note also that we use the same symbol to denote 
the superfield and its first
component.

The transformations \trana\ do not leave the sigma model
action invariant. So as in the bosonic 
case we introduce  compensating
target space transformations and then derive 
the conditions for the combine
transformation to leave the  sigma model action invariant.
In addition, we { \sl require} that the 
{\sl commutator} of two such transformations
on $q$ reproduces the conformal transformations that we have
investigated in the bosonic case up to terms
 involving the fermions $\lambda$. 
We take as a  combined transformation 
$$
\eqalign{
\delta q^i&=- \alpha c(t) \lambda^i+ \alpha X(t,q)^i{}_j \lambda^j 
\cr
\delta \lambda^i&=\alpha e(t) \partial_t q^i+ \alpha X^i(t,q)\ ,}
\eqn\csix
$$
where $\alpha$ is the anti-commuting infinitesimal
parameter, $X^i$ is a vector field in $M$ and 
$X^i{}_j$ is a tangent space rotation.
However it turns out that this is a more general class than it is
required. A direct computation of the commutator of two such
transformations on $q$ reveals that is a 
transformation of the type \btran\
provided that
we set $X^i{}_j=0$. So the  transformations 
that we shall consider  are 
$$
\eqalign{
\delta q^i&=- \alpha c(t) \lambda^i
\cr
\delta \lambda^i&=\alpha e(t) \partial_t q^i+ \alpha X^i(t,q)\ .}
\eqn\scon
$$
The commutator of two such transformations is
$$
\eqalign{
[\delta_\alpha,\delta'_\alpha]q^i&=\alpha \alpha' 
[(c'e+ce')\partial_t q^i+
c'X^i+cX'^i]
\cr
[\delta_\alpha,\delta'_\alpha]\lambda^i&=\alpha \alpha' 
[(c'e+ce')\partial_t \lambda^i
\cr &
+
(e' \partial_tc+e\partial_tc')\lambda^i+ (c'\partial_jX^i
+c\partial_j X'^i)\lambda^j]\ .}
\eqn\cseven
$$
Clearly, the transformation on $q$ induced by the commutator
is of the type that we have considered in the bosonic case.

\section{Invariance of the Action}

It remains to investigate the invariance of the action of
the $N=1$ supersymmetric sigma models under the transformations \scon. 
Before, we investigate this we introduce in addition to the
superfield $q$ another fermionic N=1 superfield\foot{The superfield $\psi$ 
 has been introduced
in the context of two-dimensional sigma models 
in [\witten] and in the context of
one-dimensional sigma models in [\coles].} $\psi$. 
The superfield $\psi$  
 can be thought of as a section of the
vector bundle $S\otimes q^* E$ over the 
superspace $\Xi$, where $S$ is 
the spin bundle over 
$\Xi$ and $E$ is a vector bundle over $M$ equipped 
with a fibre metric $h$ and a 
compatible\foot{If another non-compatible
connection is chosen, we can always redefine it and rewrite
the action in terms of a compatible one, see [\hpa].} 
 connection $\nabla$, i.e. $\nabla h=0$. The 
introduction of $\psi$ is necessary
for the addition of a scalar potential in $N=1$ sigma models.
The components of the superfield $\psi$ are
a fermion $\psi$ and  auxiliary field $F$, i.e.
$$
\psi^a=\psi^a|\qquad F^a=\nabla \psi^a|
\eqn\done
$$
where
$$
\nabla\psi^a=D\psi^a+ Dq^i B_i{}^a{}_b \psi^b
\eqn\dtwo
$$
and $B$ is a connection of $E$.

The action for the superfields $q$ and $\psi$
that we shall consider is 
$$
S=-\int dt d\theta  \big[{1\over2} g_{ij} Dq^i \partial_t q^j+ 
{1\over 3!} c_{ijk} Dq^iDq^jDq^k 
+A_i Dq^i-{1\over2} h_{ab} \psi^a\nabla \psi^b+m s_a \psi^a\big]\ ,
\eqn\supact
$$
where $g$ and $A$ are the metric and one-form coupling as the 
bosonic case, respectively,
$c_{ijk}$ is a three-form on $M$ and $s_a$ is a section of
the dual of the vector bundle $E$. The 
parameter $m$ has the
dimension of  mass.
 This is not the most general action for  $q$ and
$\psi$ (see [\coles] and appendix B). However, this choice 
will suffice for our purpose.

To compare the action that we are 
proposing above to that of the
bosonic theory, we expand \supact\ 
in components and eliminate the
auxiliary field using the field 
equations. Then  we find
$$
\eqalign{
S=\int dt &  \big[ {1\over2} g_{ij}
 \partial_t q^i \partial_t q^j
 +{1\over2} g_{ij} \lambda^i
\nabla_t^{(+)} \lambda^j-{1\over 3!}
 \partial_{[i} c_{jkm]}\lambda^i
\lambda^j\lambda^k \lambda^m
\cr
&+ A_i \partial_t q^i- {1\over2} F_{ij} \lambda^i \lambda^j+ h_{ab}
\psi^a\nabla_t\psi^b-{1\over4} G_{ij ab}
\lambda^i \lambda^j \psi^a\psi^b
\cr & - m\nabla_i s_a \lambda^i \psi^a 
-{m^2\over2} h^{ab} s_a s_b\big]\ ,}
\eqn\dthree
$$
where
$$
G_{ij}{}^a{}_b=\partial_{i} B_{j}^a{}_b-
\partial_{j} B_{i}^a{}_b+ B_i^a{}_c B_j^c{}_b-B_j^a{}_c
B_i^c{}_b
\eqn\dfour
$$
is the curvature of $B$,
$$
\nabla_t^{(+)} \lambda^i=\nabla_t \lambda^i-
 c^i{}_{jk} \partial_t q^j \lambda^k\ ;
\eqn\dfive
$$
$\nabla$ is associated to the Levi-Civita 
connection of the metric $g$.
The bosonic part of this action is that of the
 models considered in section
two with scalar potential
$$
V(q)= {m^2\over2} h^{ab} s_a s_b 
\eqn\dsix
$$
which is proportional to the length of the section $s$.

To continue it is convenient to rewrite the transformations 
\scon\ in terms of superfields
as
$$
\delta_\eta q^i= -D\eta c(t) Qq^i- 
\eta (e-c) \partial_tq^i+\eta X^i(t,q)\ ,
\eqn\suptran
$$
where $\eta=\eta(\theta)$ is a commuting 
constant infinitesimal parameter 
with components
$$
0=\eta|, \qquad \alpha= D\eta|\ .
\eqn\dseven
$$
We also take that the transformation induced 
by the odd vector fields of $\Xi$
on the $\psi$ multiplet together 
with a compensating transformation
generated by a fibre rotation $L^a{}_b(q, t)$ in $E$ as  
$$
\delta \psi^a=-D\eta \ell(t) Q\psi^a +\eta L^a{}_b(q, t) 
\psi^b\ -B_i{}^a{}_b \eta X^i(t,q),
\eqn\deight
$$ 
where $\ell(t)$ is a new function of $t$ 
which can be chosen to be
 independent from $e$ and $c$ in \suptran.
The conditions for the invariance of 
the action involving only the
superfield $q$ under \suptran\
 are
$$
\eqalign{
\nabla_{(i} X_{j)}&=-\partial_t\, c\, g_{ij}
\cr
\nabla_{[i} X_{j]}&=0
\cr
e&=c
\cr
2\partial_t X^i+
X^i F_{ij}&=0
\cr
X^i c_{ijk}&=0
\cr
{\cal L}_X c_{ijk}&=4 \partial_tc\, c_{ijk}\ ,}
\eqn\conecone
$$
where ${\cal L}_X$ is the Lie derivative with respect to $X$.
Next the conditions for the invariance
 of the part of the action involving
the superfield $\psi$ are
$$
\eqalign{
X^i(q,t) G_{ij}{}^a{}_b&=0
\cr
L_{ab}&=0
\cr
X^i \nabla_i s_a&= \partial_t c s_a
\cr
\ell&=c\ .}
\eqn\tcon
$$

The commutator of two N=1 superconformal
 transformations of the components
$\psi$ and $F$ of the superfield $\psi$ gives
$$
\eqalign{
[\delta_\alpha, \delta_\beta]\psi^a&=2 
\alpha \beta c c' \nabla_t\psi^a-B_i{}^a{}_b 
[\delta_\alpha, \delta_\beta]q^i \psi^b
\cr
[\delta_\alpha, \delta_\beta]F^a&=2
 \alpha \beta c c' \nabla_tF^a- \partial_t(c c')\alpha
\beta F^a-B_i{}^a{}_b  [\delta_\alpha, \delta_\beta]q^i F^b\ .}
\eqn\dnine
$$
\section{Special Cases}
A choice of $c$ and $X$ that leads to invariance 
of the sigma model action
 under the superconformal algebra  \supconf\ is
$$
\eqalign{
c&=c_0+t c_1
\cr
X_i(q,t)&=X_i(q)=\partial_i f(q)\ ,}
\eqn\solcon
$$
i.e. $X$ is chosen to be independent from $t$.
Substituting \solcon\ into \conecone, 
the remaining conditions become
$$
\eqalign{
\nabla_{(i} X_{j)}&=-c_1\, g_{ij}
\cr
e&=c= c_0+t c_1
\cr
X^i F_{ij}&=0
\cr
X^i c_{ijk}&=0
\cr
{\cal L}_X c_{ijk}&=4 c_1\, c_{ijk}\ .}
\eqn\fcon
$$
Similarly substituting \solcon\ in \tcon, we find
$$
\eqalign{
X^i G_{ij}{}^a{}_b&=0
\cr
L_{ab}&=0
\cr
X^i \nabla_i s_a&= c_1 s_a
\cr
\ell&=c_0+tc_1\ .}
\eqn\tcona
$$
The conditions \fcon\ and \tcona\ are those that are 
necessary and sufficient for the
supersymmetric sigma models to be invariant  under the  
transformations generated by the $Osp(1|2)$ algebra in
\supconf. 
In particular, these conditions imply that
 $X$ generates a homothetic
motion on the target space $M$, 
$X$ is expressed in terms of a  homothetic 
potential and in addition it  has 
vanishing contractions with $F$ and $G$.
The first three conditions are 
precisely the conditions that we have 
derived in the bosonic case.
It remains to derive the condition for the scalar 
potential $V$. For this,
we apply the Lie derivative on $V$  with respect 
to $X$ and use \tcona. We find that
$$
X^i \partial_iV-2 c_1 V=0
\eqn\dten
$$
in agreement with the result in the bosonic case 
after an appropriate rescaling
of $X$. In the subclass  of models without 
the fermionic multiplet $\psi$
and without a coupling to a magnetic 
field the above conditions for
$Osp(1|2)$ invariance reduce to those
of [\michb].

\section{Examples}

\subsection{Dirac Monopole}

One example of a model that admits 
$N=1$ superconformal symmetry is
that of a particle in $\bR^3$ in 
the background of a Dirac monopole
without scalar potential. It is 
straightforward to verify that in this
case all the conditions are met.
Moreover a potential can be 
added provided that it can be written
as the length of a section. To 
illustrate this, let us consider the
case of the model that it is also 
invariant under $so(3)$ rotations.
We introduce a single fermionic 
multiplet $\psi$ and choose $h_{11}=1$.
We also take $B=0$. Since the 
potential in this case is $V=|q|^{-2}$,
we can take 
$$
s=|q|^{-1}\ .
\eqn\eone
$$
Then all the conditions are satisfied
and such a model is invariant under the action
 of the $Osp(1|2)\times SO(3)$  group.

\subsection{Particles propagating in cones}

Another example can be found by 
considering spinning particles propagating
in cones, $M=C(N)$, as in the bosonic case in section (2.3). To 
demonstrate this we first
choose the metric in $M$, Maxwell field $F$ and the vector 
field that generates the homothetic
motion as in the bosonic case. Moreover, we assume 
that the three-form $c$ vanishes. It remains
to construct the scalar potential. For this, we 
consider a vector bundle $\tilde E$ on $N$ with
connection
$\tilde B$ and fibre metric
$\tilde h$. Then we pull  back $E$ on $M$ using the 
projection, $\pi\quad M=C(N)\rightarrow N$, such that
$$
\pi:\quad (r, x)\rightarrow x\ ,
\eqn\etwo
$$
where $x\in N$. This projection is well defined 
everywhere apart from the apex of
the cone at $r=0$.
For the geometric data necessary to describe the couplings of the
$\psi$ multiplet, we choose $E=\pi^*\tilde E$,
$B=\pi^*\tilde B$ and
$h=\pi^*\tilde h$. It remains to pick a 
section in the dual of $E$. For this we take
$$
s=r^{-1} \pi^*\tilde s\ ,
\eqn\ethree
$$
where $\tilde s$ is a section in the 
dual vector bundle of $\tilde E$.
It is straightforward to observe that 
a sigma model with the above couplings
satisfies all the conditions to admit 
$Osp(1|2)$ superconformal symmetry.
If in addition $N=G/H$ and the various 
other geometric data like the sigma model
metric $g$, the connection $B$, the 
fibre metric $h$ and the section $s$ are
invariant under the action of $G$, the 
model has a $G\times Osp(1|2)$ symmetry.

\chapter{Superconformal Symmetry and the N=2B Model}

\section{ Superconformal Algebra}

The methods developed to investigate the superconformal properties
of the N=1 supersymmetric sigma model can be extended to describe
the superconformal properties of the N=2B model. We 
shall give a general treatment
of the problem and then we shall specialize to find the 
conditions that are necessary for such
models to be invariant under the subgroup $SU(1,1|1)$ of the 
superconformal group of the $N=2B$ superspace. 
To describe the $su(1,1|1)$ algebra, we 
begin with the algebra $osp(1|2)$ in
 \supconf\ and relabel the generator of the 
supersymmetry transformations and the
generator of the special superconformal 
transformations by $Q_0=Q$ and by $S_0=S$,
respectively. Then we add second 
supersymmetry generator $Q_1$ and second
special superconformal generator $S_1$ 
together with the $R$ symmetry generator
of the N=2B supersymmetry algebra. The non-vanishing 
(anti-)commutators of the $su(1,1|1)$ algebra
are
$$
\eqalign{
\{Q_m,Q_n\}&=2H\delta_{mn},\qquad [Q_m,L]=Q_m,\qquad 
\{S_m,Q_n\}=L\delta_{mn}+\epsilon_{mn}R,
\cr  
\{S_m, S_n\}&=2T\delta_{mn},\qquad
[L,T]=T,\qquad [L, S_m]= 2S_m,
\cr 
[Q_m,T]&={1\over2} S_m, 
\qquad [S_m,H]=-{1\over2} Q_m, \qquad
[H,L]=H, \qquad [H,T]={1\over2} L, 
\cr 
[L,T]&=T,
\qquad
[R, Q_m]=\epsilon_{mn} Q_n, \qquad 
[R, S_m]=\epsilon_{mn} S_n\ ,}
\eqn\supconftwo
$$
where $m,n=0,1$.

\section{Invariance of the Action}

To construct  superconformal transformations 
that obey the above algebra, one should combine
the extended N=2B supersymmetry transformations 
of the one-dimensional sigma models
[\coles]
with the (2,0) supersymmetry transformations 
of the two-dimensional massive
sigma models [\hullp] and the superconformal 
transformations of the  N=1 models of the previous section.
These lead to the ansatz for the extended 
superconformal transformations
$$
\eqalign{
\delta_\eta q^i&=-D\eta c(t) I^i{}_j Dq^j+\eta Y^i(q,t)
\cr
\delta_\eta \psi^a&=-D\eta c(t) I^a{}_b \nabla\psi^b -(B_i)^a{}_b
\delta_\eta q^i \psi^b+\eta L^a{}_b \psi^b+m D\eta G^a\ ,}
\eqn\ecttwo
$$
where $\eta$ is a commuting constant infinitesimal parameter
such that
$$
\eta|=0\ ,
\eqn\efive
$$ 
$c(t)$ is a function of $\bR$,  $I^i{}_j$
is an endomorphism of the tangent bundle of the 
sigma model manifold, $I^a{}_b$ and
$L^a{}_b$
are endomorphisms\foot{A special case is 
$I^a{}_b=0$, see [\witten].}  of the 
$E$ vector bundle,
and
$G^a$ is a section of 
$E$. We have chosen the same function $c$ in the transformations
of both $q$ and $\psi$ superfields in \ecttwo\ because 
they are generated by the same odd vector field in superspace.

The conditions for the part of the action 
\supact\ that involves only the superfield
$q$ to be invariant under the 
transformations \ecttwo\ are the following:
$$
\eqalign{
g_{ik} I^k{}_j+(j,i)&=0
\cr
\nabla^{(+)}_{(i} I_{j)k}&=0
\cr
\partial_{[i} \big(I^m{}_j c_{|m|kl}\big)-
2 I^m{}_{[i} \partial_{[m}c_{jkl}&=0
\cr
\partial_tc I_{ij}+\nabla_{[i} Y_{j]}+
Y^m c_{mij}+c I^m{}_{[i} F_{j]m}&=0
\cr
Y^j F_{ji}+\partial_t Y^j g_{ji}&=0
\cr
\nabla_{(i} Y_{j)}&=0
\cr
-{1\over2} \partial_t c \nabla_{[i} I_{jk]}+ 
{1\over6}\partial_t c \nabla^{(+)}_{[i} I_{jk]}
+{1\over6} Y^m (dc)_{mijk}-c
\nabla_{[i} \big( I^m{}_j F_{k]m}\big)&=0\ .}
\eqn\esix
$$
The first three conditions are required 
from extended N=2B supersymmetry and
they have been extensively investigated 
in the literature. Part of fourth
condition is also due to the requirement 
for invariance of the action \supact\ under  extended
supersymmetry and the rest is due to the requirement for invariance
of \supact\ under extended superconformal
transformations.
Next the conditions for invariance of part 
of the action \supact\ that involves the
superfield $\psi$ under the transformations 
\ecttwo\ are as follows:
$$
\eqalign{
h_{ac} I^c{}_b+ (b,a)&=0
\cr
\nabla_i I^a{}_b&=0
\cr
I^m{}_i G_{mj}{}^a{}_b-(j,i)&=0
\cr
s_a G^a&={\rm const}
\cr
-c \nabla_js_a I^j{}_i+c \nabla_i\big(s_b I^b{}_a\big)
+\nabla_i G^b h_{ba}&=0
\cr
L^a{}_b&=0
\cr
Y^j G_{ji}{}^a{}_b&=0
\cr
Y^i \nabla_i s_a+\partial_t c s_b I^b{}_a&=0\
.}
\eqn\eseven
$$
The first four conditions and part of 
the fifth have been expected 
and arise from the
requirement of the invariance of the action 
 \supact\ under N=2B extended supersymmetry,
the rest are due to the additional requirement 
of superconformal invariance.

\section{Closure of the Algebra}

We shall first investigate the closure of 
the algebra of transformations
on $q$. A direct computation reveals that
$$
\eqalign{
[\delta_\eta, \delta_\zeta]q^i&=-D\eta D\zeta c c' 
N(I)^i{}_{jk} Dq^j Dq^k
\cr
+&2D\eta D\zeta c c' (I^2)^i{}_j \partial_t q^i
\cr
+&D\eta D\zeta  I^i{}_j(c' Y^j+c  Y'^j)
\cr
+&D\eta D\zeta (c'Dc  +cDc' )(I^2)^i{}_j Dq^j
\cr
-&\eta D\zeta c' {\cal L}_{Y}I^i{}_j Dq^j+\zeta D\eta c {\cal
L}_{Y'}I^i{}_j Dq^j\ ,}
\eqn\eeight
$$
where $N(I)$ is the Nijenhuis tensor of the 
endomorphism $I$, and $(c,Y)$ and 
$(c',Y')$ are associated with the transformations 
with parameters $\eta$ and $\zeta$,
respectively. The endomorphism $I$ 
associated with the $\eta$
transformation can be chosen to be 
different from a $I'$ associated with
 $\zeta$ transformation. However,
we have taken them to be equal $I=I'$ 
since this is a minimal choice. 

The commutator of two transformations 
\ecttwo\ on the superfield $\psi$ is
$$
\eqalign{
[\delta_\eta, \delta_\zeta]\psi^a=&-2D\eta D\zeta 
c c' (I^2)^a{}_b
\nabla_t\psi^b- B_i{}^a{}_b 
[\delta_\eta, \delta_\zeta]q^i \psi^b
\cr &
+ m D\eta  D\zeta (c' Dc +c Dc') (I^2)^a{}_b s^b
\cr &
+m \eta D\zeta Y^i 
\nabla_iG'^a-m D\eta \zeta Y'^i \nabla_iG^a
\cr &
-m D\eta D\zeta (c' I^a{}_b \nabla_i G^b
+c' I^a{}_b \nabla_i G'^b
-c I^j{}_i \nabla_jG'^a-c' I^j{}_i \nabla_jG^a) Dq^i
\cr &
-D\eta  D\zeta (c'Dc+c Dc') (I^2)^a{}_b {\cal S}^b\ ,}
\eqn\enine
$$
where
$$
{\cal S}^a=-\nabla\psi^a+h^{ab} s_b
\eqn\eten
$$
is the field equation of the $\psi$ superfield.

It remains to compute the commutator of the 
superconformal transformations
of the  N=1 model with the extended ones
 we have given above. We find that
$$
\eqalign{
[\delta_\eta, \delta_\zeta]q^i&= -\eta D\zeta c' 
{\cal L}_X I^i{}_j Dq^j 
\cr & 
+D\eta D\zeta c' Dc I^i{}_j Qq^j
+D\eta D\zeta c Qc' I^i{}_j Dq^j
\cr &
+ D\eta D\zeta (c' I^i{}_j X^j+ c Y'^i)}
\eqn\wone
$$
and
$$
\eqalign{
[\delta_\eta, \delta_\zeta]\psi^a&= D\eta D\zeta 
(c' Dc+c Dc') I^a{}_b \nabla\psi^b
\cr &
-B_i{}^a{}_b [\delta_\eta, \delta_\zeta]q^i \psi^b
\cr &
-m \eta D\zeta X^i \nabla_iG'^a\ ,}
\eqn\wtwo
$$
where $(c, X, \eta)$ are associated with the N=1 
superconformal transformations
and $(c', Y', G'^a, \zeta)$ are associated with the extended 
superconformal transformations.

\section{$SU(1,1|1)$ Invariant Models}

To find the conditions for a $N=2B$ 
supersymmetric sigma model to admit
$SU(1,1|1)$ superconformal invariance, we 
shall examine each transformation
generated by $SU(1,1|1)$ to be a symmetry separately.
The transformations generated by $Q_0$ and $S_0$
 have already been investigated in 
section (4.3).
The transformation generated by $Q_1$ is given in  \ecttwo\ for
$$
c=1,
\qquad
Y=0, \qquad
L=0\ . 
\eqn\wthree
$$
This leads to the standard extended 
supersymmetry transformations of [\coles].
The conditions for the invariance of the action are
$$
\eqalign{
g_{ik} I^k{}_j+(j,i)&=0
\cr
\nabla^{(+)}_{(i} I_{j)k}&=0
\cr
\partial_{[i} \big(I^m{}_j c_{|m|kl}\big)-
2 I^m{}_{[i} \partial_{[m}c_{jkl}&=0
\cr
 I^m{}_{[i} F_{j]m}&=0
 \cr
 h_{ac} I^c{}_b+ (b,a)&=0
\cr
\nabla_i I^a{}_b&=0
\cr
I^m{}_i G_{mj}{}^a{}_b-(j,i)&=0
\cr
s_a G^a&={\rm const}
\cr
-\nabla_js_a I^j{}_i+\nabla_i\big(s_b I^b{}_a\big)
+\nabla_i G^b h_{ba}&=0\ .}
\eqn\wfour
$$
In addition, the algebra of two extended 
supersymmetries closes to translations
 provided that
$$
\eqalign{
N(I)&=0
\cr
I^i{}_j I^j{}_k&=-\delta^i{}_k
\cr
I^a{}_b I^b{}_c&=-\delta^a{}_c
\cr
I^a{}_b \nabla_iG^b-I^j{}_i \nabla_jG^a&=0\ .}
\eqn\wfive
$$
Most of the conditions in \wfour\ and \wfive\ are 
well known and have been derived
in the context of one-dimensional sigma models 
in [\coles, \sfetsos, \gibbonsp].
In particular they imply that the target space 
is a hermitian manifold\foot{Under certain
conditions, the integrability of the complex structure
can be lifted; see [\vn,\hpb, \gpb, \hull].} equipped 
with a holomorphic vector bundle
$E$. The line bundle associated to the magnetic coupling
is also holomorphic. The commutator of extended 
supersymmetry generated by $Q_1$ with 
that of  N=1 supersymmetry
generated by $Q_0$ vanishes without any further conditions. 
Next the commutator of
extended supersymmetry transformation generated by $Q_1$ with 
that of special superconformal 
transformation
$$
\eqalign{
\delta_\eta q^i&= -D\eta t Qq^i+\eta X^i(q)
\cr
\delta_\eta \psi^a&= -D\eta t Q\psi^a}
\eqn\wsix
$$
generated by $S_0$ is
$$
\eqalign{
[\delta_\eta, \delta_\zeta]q^i&=\alpha \beta I^i{}_j X^j
\cr
[\delta_\eta, \delta_\zeta]\lambda^i&=-\alpha \beta
 \big[{\cal L}_XI^i{}_j+ I^i{}_j] \lambda^i + 
 \alpha \beta \partial_k(I^i{}_j X^j) \lambda^k
 \cr
[\delta_\eta, \delta_\zeta]\psi^a&=-B_i{}^a{}_b 
[\delta_\eta, \delta_\zeta]q^i \psi^b
\cr
[\delta_\eta, \delta_\zeta]F^a&=-\alpha \beta I^a{}_b F^b-
\alpha \beta X^i \nabla_iG^a+\alpha \beta G_{ij}{}^a{}_b I^i{}_k X^k
 \lambda^j \psi^b
-B_i{}^a{}_b [\delta_\eta, \delta_\zeta]q^i \psi^b\ ,}
\eqn\wseven
$$
where $\alpha=D\eta|$ and $\beta=D\zeta|$.
The above commutator should close to the 
transformation generated by the $R$
symmetry of the $N=2B$ superalgebra. This is 
the case provided that
$$
\eqalign{
{\cal L}_XI^i{}_j&=0 
\cr
X^i\nabla_i G^a&=0
\cr
G_{ij}{}^a{}_b I^i{}_k X^k&=0\ .}
\eqn\weight
$$
The latter condition is not independent condition but it is 
implied by the fact that
$G_{ij}{}^a{}_b$ is (1,1) form with respect to $I$ and that
 the contraction of $X$ with 
$G_{ij}{}^a{}_b$ vanishes.

Next let us consider the case of the special
 superconformal transformation
generated by $S_1$. For this
we take
$$
\eqalign{
c&=t
\cr
Y(t,q)&=Y(q)\ .}
\eqn\wnine
$$
The additional conditions required for invariance of 
the action are as follows:
$$
\eqalign{
I_{ij}+\nabla_{[i} Y_{j]}+Y^m c_{mij}&=0
\cr
Y^j F_{ji}&=0
\cr
\nabla_{(i} Y_{j)}&=0
\cr
-{1\over2}  \nabla_{[i} I_{jk]}+ 
{1\over6} \nabla^{(+)}_{[i} I_{jk]}+{1\over6} Y^m (dc)_{mijk}&=0
\cr
\nabla_iG^a &=0
\cr
Y^j G_{ji}{}^a{}_b&=0
\cr
Y^i \nabla_i s_a+ s_b I^b{}_a&=0\ . }
\eqn\wten
$$
Observe that $G^a$ is parallel. This leads to a 
simplification of the equations
\wfour, \wfive\ and \weight. From the last 

equation in \wten\ together with
the hermiticity of the fibre metric $h$, we find that
the scalar potential is {\sl invariant} 
under the isometry $Y$.

Some of the  equations in \wfour\ and 
in \wten\ can be expressed in different
ways. In fact they can be further 
simplified by choosing as 
$\nabla^{(+)}$ the unique metric connection
of the hermitian geometry with antisymmetric torsion 
such that $\nabla^{(+)}I=0$, i.e  $M$ is a KT manifold.

Now first us consider the commutator 
of two transformations generated
by $S_1$. This commutator closes to
a transformation generated by $T$
provided that
$$
\eqalign{
X^i&= I^i{}_j Y^j
\cr
{\cal L}_YI^i{}_j&=0\ .}
\eqn\weleven
$$

The rest of the commutators are 
satisfied without further conditions.
In the subclass  of models 
without the fermionic multiplet $\psi$
and without a coupling to a
 magnetic field the above conditions for
$SU(1,1|1)$ invariance are related  to those
of [\michb].

To summarize,  $SU(1,1|1)$ superconformal invariance 
of a sigma model with action  \supact\
 requires that
the sigma model manifold $M$ is hermitian and the
 vector bundle $E$ is holomorphic
with compatible fibre metric and complex structure. 
In addition $E$ admits a parallel
section (which can be chosen to vanish) and 
a holomorphic section $s$.
Moreover $M$ admits a holomorphic homothetic
 motion generated by $X$ and a holomorphic
isometry generated by $Y$ which are related as in \weleven.
The scalar potential scales under the 
homothetic motion with weight $-2$ and
it is invariant under the isometry.

\chapter{Superconformal Symmetry and the N=4B Model}

\section{Superconformal Algebra}

The investigation of the superconformal symmetries of the $N=4B$
supersymmetric sigma model can be done in a way similar to that of the
$N=2B$ supersymmetric sigma 
model in the previous sections.
We shall begin with a general treatment of the problem 
and then we shall specialize to find the 
conditions that are necessary for such
models to be invariant under the 
subgroup $D(2,1|\alpha)$ of the 
superconformal group of the $N=4B$ superspace. 
The non-vanishing (anti)-commutators of the  $D(2,1|\alpha)$
 superconformal subalgebra
are as follows:
$$
\eqalign{
\{Q_m,Q_n\}&=2H\delta_{mn},\qquad [Q_m,L]=Q_m,
\cr  
\{S_m, S_n\}&=2T\delta_{mn},\qquad
[L,T]=T,\qquad [L, S_m]= 2S_m,
\cr 
[Q_m,T]&={1\over2} S_m, 
\qquad [S_m,H]=-{1\over2} Q_m, \qquad
[H,L]=H, \qquad [H,T]={1\over2} L, 
\cr 
[L,T]&=T,
\qquad
[R^r_\pm, Q_m]=(t^r_\pm)_{mn} Q_n, 
\qquad [R^r_\pm, S_m]=(t^r_\pm)_{mn} S_n
\cr
\{S_m,Q_n\}&=L\delta_{mn}+{\alpha\over 1+\alpha} 
(t^r_+)_{mn} R^r_++{1\over 1+\alpha}
(t^r_-)_{mn} R^r_-
\cr
[R^r_\pm, R^s_\pm]&=\pm 2 \epsilon^{rst} R^t_\pm
\ ,}
\eqn\supconffour
$$
where $\{S_m; m=0,\dots, 3\}$ are the generators of the special
superconformal transformations, 
$\{Q_m; m=0,\dots, 3\}$ are the supersymmetry
generators, $\{R^r_\pm; r=1,2,3\}$ are the generators of 
the $SO(4)$ R-symmetry of the
$N=4B$ supersymmetry algebra and
$$
(t^r_\pm)_{mn}=2 \delta^0{}_{[m} \delta^r{}_{n]}\pm 
\epsilon^{0r}{}_{mn}
\eqn\blackn
$$
is a basis of self-dual and anti-self-dual two forms in $\bR^4$.
It remains to
investigate the conditions under which the action
\supact\ admits such extended superconformal symmetry. For this,
we consider the transformations
$$
\eqalign{
\delta_\eta q^i &=-D\eta^r c_r(t) (I_r)^i{}_j Dq^j+\eta^r Y_r^i(q,t)
\cr
\delta_\eta \psi^a &=-D\eta^r c_r(t) (I_r)^a{}_b \nabla\psi^b 
-(B_i)^a{}_b \delta_\eta q^i \psi^b
+\eta^r (L_r)^a{}_b \psi^b+m D\eta^r G_r^a 
\ ,}
\eqn\ect
$$
where $\{c_r; r=1,2,3\}$ are functions on $\bR$,  $\{\eta^r; r=1,2,3\}$ 
are a commuting constant
infinitesimal parameters such that $\eta_r|=0$,
$\{(I_r)^i{}_j; r=1,2,3\}$ are endomorphisms of the tangent 
bundle of the sigma model manifold,
$\{(I_r)^a{}_b; r=1,2,3\}$ and
$\{(L_r)^a{}_b; r=1,2,3\}$
are endomorphisms of the $E$ vector bundle and
 $\{G_r^a; r=1,2,3\}$ are sections of 
$E$.

\section{Invariance of the Action}
The conditions for the part of the action \supact\ that 
involves only the superfield
$q$ to be invariant under the 
transformations \ect\ are the following:
$$
\eqalign{
g_{ik} (I_r)^k{}_j+(j,i)&=0
\cr
\nabla^{(+)}_{(i} (I_r)_{j)k}&=0
\cr
\partial_{[i} \big((I_r)^m{}_j c_{|m|kl}\big)
-2 (I_r)^m{}_{[i} \partial_{[m}c_{jkl}&=0
\cr
\partial_tc_r (I_r)_{ij}+\nabla_{[i} (Y_r)_{j]}+Y_r^m c_{mij}
+c_r (I_r)^m{}_{[i} F_{j]m}&=0
\cr
Y_r^j F_{ji}+\partial_t Y^j g_{ji}&=0
\cr
\nabla_{(i} (Y_r)_{j)}&=0
\cr
-{1\over2} \partial_t e \nabla_{[i} (I_r)_{jk]}+ 
{1\over6}\partial_t e \nabla^{(+)}_{[i} (I_r)_{jk]}
+{1\over6} Y^m_r (dc)_{mijk}-e
\nabla_{[i} \big( (I_r)^m{}_j F_{k]m}\big)&=0\ .}
\eqn\zone
$$
The first three conditions are required from 
extended supersymmetry and
they have been extensively investigated in the literature. Part of fourth
condition is also due to the requirement for invariance of
 the action \supact\ under  extended
supersymmetry and the rest is due to
 the requirement for invariance
of \supact\ under extended superconformal
transformations.
Next the conditions for invariance of 
part of the action \supact\ that involves the
superfield $\psi$ under the 
transformations \ect\ are as follows:
$$
\eqalign{
h_{ac} (I_r)^c{}_b+ (b,a)&=0
\cr
\nabla_i (I_r)^a{}_b&=0
\cr
(I_r)^m{}_i G_{mj}{}^a{}_b-(j,i)&=0
\cr
s_a G^a_r&={\rm const}
\cr
-c_r \nabla_js_a (I_r)^j{}_i+
c_r \nabla_i\big(s_b (I_r)^b{}_a\big)
+\nabla_i G_r^b h_{ba}&=0
\cr
L_r^a{}_b&=0
\cr
Y_r^j G_{ji}{}^a{}_b&=0
\cr
Y_r^i \nabla_i s_a+\partial_t c_r s_b (I_r)^b{}_a&=0\
.}
\eqn\ztwo
$$
The first four conditions and part of the fifth 
have been expected and arise from the
requirement of the invariance of the action  
\supact\ under extended supersymmetry,
the rest are due to the additional 
requirement of superconformal invariance.

\section{Closure of the Algebra}

We shall first investigate the closure 
of the algebra of transformations
on $q$. A direct computation reveals that
$$
\eqalign{
[\delta_\eta, \delta_\zeta]q^i&=-D\eta_rD\zeta_s c_r c_s 
N(I_r, I_s)^i{}_{jk} Dq^j Dq^k
\cr
+&2D\eta_rD\zeta_s c_r c_s (I_{(r} I_{s)})^i{}_j \partial_t q^i
\cr
+&D\eta_rD\zeta_s (c_s (I_s)^i{}_j Y_r^j
+c_r (I_r)^j{}_i Y_s^j)
\cr
+&D\eta_rD\zeta_s (c_sDc_r (I_sI_r)^i{}_j 
+c_rDc_s (I_rI_s)^i{}_j)Dq^j
\cr
-&\eta_rD\zeta_s c_s 
{\cal L}_{Y_r}(I_s)^i{}_j Dq^j+\zeta_sD\eta_r c_r {\cal
L}_{Y_s}(I_r)^i{}_j Dq^j\ ,}
\eqn\zthree
$$
where $N(I_r, I_s)$ is the Nijenhuis tensor of 
the endomorphisms $I_r$ and $I_s$.

The commutator of two transformations
 on the superfield $\psi$ is
$$
\eqalign{
[\delta_\eta, \delta_\zeta]\psi^a=&-2D\eta_r D\zeta_s c_r c_s 
(I_{(r} I_{s)})^a{}_b
\nabla_t\psi^b- B_i{}^a{}_b 
[\delta_\eta, \delta_\zeta]q^i \psi^b
\cr &
+ m D\eta_r  D\zeta_s c_s Dc_r (I_s I_r)^a{}_b s^b
+m D\eta_r  D\zeta_s c_r Dc_s (I_r I_s)^a{}_b
s^b
\cr &
+m \eta_r D\zeta_s Y^i_r \nabla_iG_s^a
-m D\eta_r \zeta_s Y^i_s \nabla_iG_r^a
\cr &
-m D\eta_r D\zeta_s (c_s (I_s)^a{}_b \nabla_i G_r^b
+c_r (I_r)^a{}_b \nabla_i G_s^b
\cr &
-c_r (I_r)^j{}_i \nabla_jG_s^a
-c_s (I_s)^j{}_i \nabla_jG_r^a) Dq^i
\cr &
-D\eta_r  D\zeta_s c_s Dc_r (I_s I_r)^a{}_b {\cal S}^b
-D\eta_r  D\zeta_s c_r Dc_s (I_r
I_s)^a{}_b {\cal S}^b\ ,}
\eqn\zfour
$$
where
$$
{\cal S}^a=-\nabla\psi^a+h^{ab} s_b
\eqn\zfive
$$
is the field equation of the $\psi$ superfield.

The commutator of \ect\ transformations with those 
of N=1 superconformal symmetries
of section (4.1) is
$$
\eqalign{
[\delta_\eta, \delta_\zeta]q^i&= -\eta D\zeta c_r' 
{\cal L}_X I_r^i{}_j Dq^j 
\cr & 
+D\eta D\zeta_r c_r' Dc I_r^i{}_j Qq^j
+D\eta D\zeta_r c Qc_r' I_r^i{}_j Dq^j
\cr &
+ D\eta D\zeta_r (c_r' I_r^i{}_j X^j+ c Y_r'^i)}
\eqn\zsix
$$
and
$$
\eqalign{
[\delta_\eta, \delta_\zeta]\psi^a&= D\eta D\zeta_r
 (c_r' Dc+c Dc_r') I_r^a{}_b \nabla\psi^b
\cr &
-B_i{}^a{}_b [\delta_\eta, \delta_\zeta]q^i \psi^b
\cr &
-m \eta D\zeta_r X^i \nabla_iG_r'^a\ ,}
\eqn\zseven
$$
where $(c, X, \eta)$ are associated with the 
N=1 superconformal transformations
and $(c', Y_r', G_r'^a, \zeta)$ 
are associated with the \ect\
superconformal transformations.

\section{$D(2,1|\alpha)$-invariant models}

To investigate the conditions for the
 $N=4B$ supersymmetric sigma models
to admit a $D(2,1|\alpha)$ superconformal 
symmetry, we shall examine the
conditions required for each 
transformation generated by $D(2,1|\alpha)$
to be a symmetry, separately. The conditions required 
for the invariance of the sigma models
under the transformations 
generated by $Q_0$ and $S_0$ have already been
investigated in section (4.3).
The extended supersymmetry 
transformations generated by $Q_r$ are given by  \ect\
for 
$$
c_r=1, \qquad Y_r=0, \qquad L_r=0\ .
\eqn\zeight
$$
The conditions for the invariance of the action are then
$$
\eqalign{
g_{ik} (I_r)^k{}_j+(j,i)&=0
\cr
\nabla^{(+)}_{(i} (I_r)_{j)k}&=0
\cr
\partial_{[i} \big((I_r)^m{}_j c_{|m|kl}\big)-2
 (I_r)^m{}_{[i} \partial_{[m}c_{jkl}&=0
\cr
(I_r)^m{}_{[i} F_{j]m}&=0
\cr
h_{ac} (I_r)^c{}_b+ (b,a)&=0
\cr
\nabla_i (I_r)^a{}_b&=0
\cr
(I_r)^m{}_i G_{mj}{}^a{}_b-(j,i)&=0
\cr
s_a G^a_r&={\rm const}
\cr
-\nabla_js_a (I_r)^j{}_i+\nabla_i\big(s_b (I_r)^b{}_a\big)
+\nabla_i G_r^b h_{ba}&=0
\
.}
\eqn\znine
$$
The conditions for the closure of the algebra of the
extended supersymmetry transformations are
that
$$
\eqalign{
N(I_r, I_s)&=0
\cr
I_{(r} I_{s)}&= -\delta_{rs}
\cr
I_{(r} I_{s)}&= -\delta_{rs}
\cr
(I_r)^a{}_b\nabla_i G^b_r-
(I_r)^j{}_i\nabla_j G^b_r+(s,r)&=0\ .}
\eqn\zten
$$
Most of the conditions \znine\ and 
\zten\ also appear in the
case of $N=4B$ supersymmetric sigma models 
without scalar potential in 
[\coles, \gibbonsp].
In what follows we shall choose the 
complex structures to obey the somewhat
stronger condition
$$
I_r I_s=-\delta_{rs} +\epsilon_{rst} I_t\ .
\eqn\zeleven
$$
It was shown in [\michb] that \zeleven\ 
and the first three conditions
in \znine\ imply that 
$$
\nabla^{(+)}_i(I_r)^j{}_k=0\ .
\eqn\ztwelve
$$
So the sigma model manifold admits a 
weak HKT structure [\twist]. Moreover, the
fourth and seventh conditions in 
 \znine\ imply that both the line bundle
associated with the magnetic field 
and the vector bundle $E$ are holomorhic
with respect to all three complex structures $\{I_r\}$.

For the extended special 
superconformal transformations generated by $S_r$,
 we choose
in \ect
$$
c_r=t\ ,\qquad Y_r(q,t)=Y_r(q)\ .
\eqn\zzone
$$
The additional conditions for the invariance of the action under such
transformations are as follows:
$$
\eqalign{
I_{rij}+\nabla_{[i} (Y_r)_{j]}+Y_r^m c_{mij}&=0
\cr
Y_r^j F_{ji}&=0
\cr
\nabla_{(i}(Y_r)_{j)}&=0
\cr
-{1\over2}  \nabla_{[i} (I_r)_{jk]}
+{1\over6} Y_r^m (dc)_{mijk}&=0
\cr
\nabla_iG^a_r&=0
\cr
Y_r^j G_{ji}{}^a{}_b&=0
\cr
Y_r^i \nabla_is_a+ s_b (I_r)^b{}_a&=0\ . }
\eqn\zzthree
$$
Observe that the above conditions imply
 that $G_r^a$ are parallel. Using this,
one can simplify some of the  
conditions in \znine\ and \zten.  In what follows
 we shall use that
$G_r^a$ are parallel. Some of the 
conditions in \zzthree\ can be expressed
in a different way, e.g. the fourth condition using \ztwelve.

The commutator of two extended special 
superconformal transformations generated
by $S_r$ closes to special 
conformal transformations provided that
$$
\eqalign{
X^i&=(I_r)^i{}_j Y_r^j
\cr
{\cal L}_{Y_r} (I_s)^i{}_j+ (s,r)&=0
\ .}
\eqn\zzfive
$$

To find the transformations associated with the 
$R\pm$ generators of $D(2,1|\alpha)$, we shall
next compute the commutator of the special 
superconformal transformations generated by $S_n$
and the supersymmetry transformations
 generated by $Q_m$. Let us begin with the
commutator of extended supersymmetry transformations 
generated by $Q_s$ with the extended
special superconformal transformations 
generated by $S_r$. We find that
$$
\eqalign{
[\delta_\alpha, \delta_\beta]q^i=&-2\alpha_r\beta_s\delta_{rs}  t
\partial_tq^i+\alpha_r \beta_s \delta_{rs} X^i -
\alpha_r \beta_s \epsilon_{rst} (I_t)^i{}_jX^j)
\cr
[\delta_\alpha, \delta_\beta]\lambda^i=&-2\alpha_r \beta_s \delta_{rs}  t
\partial_t\lambda^i+ \alpha_r \beta_s \delta_{rs}
-\epsilon_{rst} (I_t)^i{}_j \lambda^j
\cr
&+
\alpha_r \beta_s \delta_{rs} \partial_jX^i \lambda^j-
\epsilon_{rst}  \partial_k\big((I_t)^i{}_j)X^j\big)
\lambda^k
\cr
& -
\alpha_r \beta_s{\cal L}_{Y_s} (I_r)^i{}_j \lambda^j
\cr
[\delta_\alpha, \delta_\beta]\psi^a=&2\alpha_r \beta_s 
\delta_{rs} t 
\nabla_t \psi^a-
B_i{}^a{}_b [\delta_\alpha, \delta_\beta]q^i \psi^b
\cr
[\delta_\alpha, \delta_\beta]F^a=&2\alpha_r \beta_s 
\delta_{rs} t 
\nabla_t F^a-2 \alpha_r
\beta_s (I_r I_s)^a{}_b F^b-
B_i{}^a{}_b [\delta_\alpha, \delta_\beta]q^i F^b
\ ,}
\eqn\zzsix
$$
where
$\alpha_r=D\eta_r|$ and $\beta_s=D\zeta_s|$ are the
 infinitesimal parameters of the
supersymmetry and special 
superconformal transformations, respectively.
The commutator of the first supersymmetry transformation 
generated by $Q_0$ with the extended
superconformal ones generated by $S_r$ is
$$
\eqalign{
[\delta_\alpha, \delta_\beta]q^i=&-\alpha\beta_s 
(I_s)^i{}_j X^j 
\cr
[\delta_\alpha, \delta_\beta]\lambda^i
=&\alpha \beta_s (I_s)^i{}_j \lambda^j
\cr
& -\alpha \beta_s \partial_k \big((I_s)^i{}_j X^j\big) \lambda^k
\cr
[\delta_\alpha, \delta_\beta]\psi^a=&-
B_i{}^a{}_b [\delta_\alpha, \delta_\beta]q^i \psi^b
\cr
[\delta_\alpha, \delta_\beta]F^a=&\alpha \beta_s 
(I_s)^a{}_b F^b-
B_i{}^a{}_b [\delta_\alpha, \delta_\beta]q^i F^b
\ ,}
\eqn\zzeight
$$
The commutator of the special superconformal 
transformation generated by $S_0$
with extended supersymmetry ones generated by $Q_r$ is
$$
\eqalign{
[\delta_\alpha, \delta_\beta]q^i=&\alpha\beta_r (I_r)^i{}_j X^j
\cr
[\delta_\alpha, \delta_\beta]\lambda^i=&-\alpha \beta_r 
{\cal L}_X (I_r)^i{}_j \lambda^j
-\alpha \beta_r (I_r)^i{}_j \lambda^j
\cr+
& \alpha \beta_r \partial_k\big((I_r)^i{}_j X^j\big) \lambda^k
\cr
[\delta_\alpha, \delta_\beta]\psi^a=&-
B_i{}^a{}_b [\delta_\alpha, \delta_\beta]q^i \psi^b
\cr
[\delta_\alpha, \delta_\beta]F^a=&-\alpha \beta_r 
(I_r)^a{}_b F^b-
B_i{}^a{}_b [\delta_\alpha, \delta_\beta]q^i F^b
\ .}
\eqn\zznine
$$
In order the part of the anti-commutator 
$\{S_m, Q_n\}$ involving the 
$R$ generators to
be antisymmetric, we take 
$$
{\cal L}_X I_r=0\ .
\eqn\zzten
$$
Next following closely the analysis in [\michb], we choose
$$
{\cal L}_{Y_s} I_r= h \epsilon_{rst} I_t\ ,
\eqn\zzeleven
$$
where $h$ is a real number.
Then the part of the commutators on $\lambda$ 
involving the R-symmetries 
can be written
schematically as
$$
[\delta_\alpha, \delta_\beta]\lambda^i\equiv 
[\alpha^m Q_m, \beta^nS_n]\lambda^i=\alpha_m\beta_n
\delta_{mn} L\lambda^i+
\alpha_m\beta_n t^r_{mn} R_r\lambda^i\ ,
\eqn\zztwelve
$$
where
$$
\eqalign{
t^r_{0s}&=\delta^r{}_s
\cr
t^r_{s0}&=-\delta^r{}_s
\cr
t^r_{st}&=-(1+h) \epsilon_{rst}\ .}
\eqn\zaone
$$
Rewriting $t^r$ in terms of the self-dual $(t^r_+)$ 
and anti-self-dual  $(t^r_-)$
basis of
two-forms, we find that
$$
t^r=-{h\over2} (t^r_+) +{2+h\over2}(t^r_-)\ .
\eqn\zatwo
$$ 
Comparing this with the superconformal algebra 
$D(2,1|\alpha)$ in \supconffour, we find
that
$$
\alpha=-{h\over 2+h}
\eqn\zathree
$$
Finally we observe that $q$, $\psi$ and $F$ transform 
only under the $R_+$ generators
of $D(2,1|\alpha)$
while $\lambda$ transforms under both  
 $R_+$ and $R_-$ generators
of $D(2,1|\alpha)$. 
The rest of the commutators are 
satisfied without further conditions.
In the case of models without 
fermionic superfields $\psi$ and a coupling
to a magnetic field $A$, the above
 conditions for $D(2,1|\alpha)$ are
related to those derived in [\michb].

To summarize,  $D(2,1|\alpha)$ superconformal 
invariance of a sigma model
with action  \supact\ requires that
the sigma model manifold $M$ is HKT and the
 vector bundle $E$ is tri-holomorphic
with compatible fibre metric and hypercomplex 
structure. In addition $E$ admits three parallel
sections (which can be chosen to vanish) 
and a tri-holomorphic section $s$.
Moreover $M$ admits a holomorphic homothetic 
motion generated by $X$ and three holomorphic
isometries generated by $Y_r$ which are related as in \zzfive.
The scalar potential scales under the 
homothetic motion with weight $-2$ and
it is invariant under the three isometries.

\chapter{Conclusions}

We have given the conditions for a spinning particle  coupled
to a magnetic field and with a scalar potential to be invariant
under conformal and superconformal symmetries.
In particular, we have presented the conditions for such models
to admit $SL(2,\bR)$, $Osp(1|2)$, $SU(1,1|1)$  or $D(2,1|\alpha)$
(super)conformal symmetries.
We have given several examples of such systems that include
a particle propagating in a Dirac monopole background and a
particle propagating in a conical spacetime.
Moreover we have found an  interpretation of the
conformal transformations using the geometry of bundles.

In the superconformal case, we considered a special class
of one-dimensional sigma model Lagrangians with a scalar potential.
There are many other possibilities to explore. It is already known
that some of these cases have applications in the context of the
moduli spaces of four-dimensional supersymmetric black holes [\mss].
Therefore it would be of interest to explore the superconformal properties
of these more general models described by the Lagrangian of appendix B.
We leave this work for the future.

\vskip 1cm
\noindent{\bf Acknowledgments:} I would like to thank A. 
Strominger for many helpful
discussions. I am
supported by a University Research 
Fellowship from the Royal Society.

\Appendix {A}
\centerline{{ \bf  The geometry of conformal transformations}}

There is a geometric interpretation of the 
conformal transformations we are considering
in the bosonic spinning particle. To see this, 
we consider  $P=\bR\times M$ as a bundle over
$\bR$ with respect to the obvious 
projection $\pi_{{}_{\bR}}$, where $\bR$
 is the worldline and $M$
is the sigma model target space.
Let $\sigma$ be a section of $P$, then the 
sigma model maps $q$ can be written as
$$
q=\pi_{{}_{M}} \sigma
\eqn\aaone
$$
where $\pi_{M}$ is the obvious projection of $P$ onto $M$.
Next consider the  bundle isomorphisms $\alpha$ of $P$,
$$
\alpha (t, x)= (f(t), g(t,x))\ ,
\eqn\aatwo
$$
where $t\in \bR$, $x\in M$,  $f$ is a diffeomorphism of $\bR$ and $g$ is a 
diffeomorphism of $M$
for every $t$.
The bundle isomorphism $\alpha$ induces a 
transformation $U_\alpha$ on the space of sections of $P$ as
 follows:
 $$
 (U_\alpha \sigma)(x)=\alpha\big(\sigma(f^{-1}(x))\big)\ .
 \eqn\aathree
 $$
 Observe that $U_\alpha$ takes a section 
$\sigma$ at the point $x$ to another
 section $U_\alpha \sigma$ at the same point, i.e. it 
shifts the sections along the
 fibre $M$. 
 Composing $U_\alpha$ with the projection on $M$, we 
induce a transformation on the
 sigma model maps which we also denote by $U_\alpha$, i.e.
 $$
(U_\alpha(q))(x)=\pi_{{}_M} (U_\alpha \sigma)(x)\ .
\eqn\aafour
$$

Next consider an one-parameter family $\alpha_s$ of  
isomorphisms such that
$$
\alpha_{s=0}=Id_P\ .
\eqn\aafive
$$
The vector fields on $P$ that generate such one-parameter
 families of isomorphisms
are
$$
A=a(t) \partial_t+ X^i(t, x)\partial_i\ .
\eqn\aasix
$$
Moreover a direct computation reveals that the vector 
fields on $P$ which are
generated by the associated transformations 
$U_{\alpha_s}$ on the sections
of $P$ are
$$
\tilde A=-a(t)\partial_t \big(\pi_{{}_{\bR}}\sigma\big)^i \partial_i
+X^i(t, \pi_{{}_{\bR}}\sigma)\partial_i
\eqn\aaseven
$$
or on the sigma model maps $q$
$$
\tilde A=-a(t)\partial_tq^i \partial_i
+X^i(t, q)\partial_i\ .
\eqn\vvvv
$$
Clearly $\tilde A$ generates the 
infinitesimal transformations that we 
have investigated in section two on the sigma model maps $q$.

\Appendix {B}
\centerline{\bf{The $N=1$ Supersymmetric Action}}

The most general $N=1$ spinning particle action coupled to a magnetic
filed and with the scalar potential is
$$
\eqalign{
S=-\int dt d\theta  &\big[{1\over2} 
g_{ij} Dq^i \partial_t q^j+ 
{1\over 3!} c_{ijk} Dq^iDq^jDq^k 
-{1\over2} h_{ab} \psi^a\nabla \psi^b
\cr &
+{1\over2}m_{iab} Dq^i \psi^a \psi^b
+{1\over2} n_{ija} Dq^i Dq^j \psi^a +
{1\over3!} l_{abc} \psi^a \psi^b\psi^c
\cr & 
+f_{ia}\partial_t \psi^a+A_i Dq^i+m s_a \psi^a\big]\ ,}
\eqn\hahb
$$
where $q$ is a bosonic and $\psi$ is a fermionic $N=1$
superfield, respectively.
This action apart from the last two
 terms has been proposed in [\coles].
The term involving $A$ is a standard 
supersymmetric extension of the coupling
of a charged particle to a magnetic field. 
The last term involves the addition
of a scalar potential in the action. 
Such a term has been first proposed
in the context of two-dimensional 
sigma models in [\hullp] and it can be
easily adapted to this case. It is 
worth pointing out that the addition
of the scalar potential necessarily 
involves the  superfield $\psi$.

\refout

\bye